\documentclass[graphicx, 12pt]{article}
\usepackage{indentfirst}

\usepackage{graphicx}
\usepackage{caption}
\usepackage{subfigure}

\begin{document}

\small
\hoffset=-1truecm
\voffset=-2truecm
\title{\bf The Horava-Lifshitz Modifications of the Casimir effect
at finite temperature revisted}

\author{Hongbo Cheng\footnote
{E-mail address: hbcheng@ecust.edu.cn}\\
Department of Physics,\\ East China University of Science and
Technology,\\ Shanghai 200237, China\\
The Shanghai Key Laboratory of Astrophysics,\\Shanghai 200234,
China}

\date{}
\maketitle

\begin{abstract}
We proceed with the study of the Casimir force for parallel plates
at finite temperature in the Horava-Lifshitz (HL) theory. We find
that the HL exponent can not be chosen as an integer, or the
Casimir energy will be a constant and further the Casimir force
between two parallel plates will vanish. The higher temperature
makes the attractive Casimir force weaker, which is consistent
with the original consequence confirmed theoretically and
experimentally. We can select the HL factor adequately to lead the
thermally revised Casimir force to be similar to the standard
results for the parallel plates.
\end{abstract}

\vspace{6cm} \hspace{0cm} PACS number(s): 03.70.+k, 11.10.Ef

\noindent \textbf{I.\hspace{0.4cm}Introduction}

Under the background fields or within the quantization volume, the
Casimir effect as a direct consequence of quantum field theory
subject to a change in the spectrum of vacuum oscillations was put
forward more than seventy years ago [1]. The Casimir effect shows
a shift of the vacuum energy because of the nontrivial boundary
conditions or the topology of the space [1-3]. Afterwards the
repulsive Casimir force for a conducting spherical shell was found
by Boyer [4]. The nature of Casimir force depends on the
circumstances mentioned above. The attractive Casimir force
between two parallel plates was varied experimentally by Sparnaay
[5]. Further the precision of the measurements of Casimir effect
has been greatly improved experimentally and the more precise
measurements have been performed [5-11]. The characters of Casimir
effect including the sign of the Casimir energy and the nature of
the Casimir force have been discussed in many subjects because the
Casimir effect has something to do with various factors which
cover the boundaries, the structure of spacetime, the
topologically non-trivial backgrounds, some types of quantum field
theory and the temperature [8, 12-15]. A lot of effort has been
paid to the problems and related topics. The Casimir force can be
changed by geometry of boundaries [16-19]. The presence of extra
dimensions with their size and shape also affect the Casimir
effect [20-34]. The quantum field theory at finite temperature
shares many effects [35, 36]. The thermal influence certainly
modifies the Casimir effect and can not be neglected [37-49].

As a kind of modified field theory, the Haorava-Lifshitz (HL)
theory tries to help the gravity to become power-counting
renormalizable. The HL model proceeds the scaling of coordinates
like $t\longrightarrow b^{z}t$ and $x^{i}\longrightarrow bx^{i}$
to lead the spacetime anisotropy, where $b$ is a length constant
and $z$ is the critical exponent [50-52]. According to the energy
scale, we can adjust the value of $z$ to restore the
renormalizability of the theory [50-52]. The factor $z$ tends to
be unity at low energies and is more than unity within the
micrometer-sized region [50-52]. The HL approach was used to
describe black holes, gravitational waves etc. to show its
influence [53-57]. The kinetic terms with higher order spatial
derivatives appear in the Lagrangian due to rescaling the
coordinates in the context of HL issue [50-52, 58]. These new
kinetic terms recover the renormalization of modified model while
break the Lorentz symmetry [48-50, 56]. The Lorentz violation may
happen at high energies or in the extremely-small region [52]. In
the cosmology, the finite action from the Horava-Lifshitz theory
leads the beginning of the flat and homogeneous universe while the
different versions of the issue can result in different
cosmological models [58]. The Horava-Lifshitz gravity only involve
the regular black holes [58]. The Horava-Lifshitz gravity is
thought as a candidate of quantum gravity. The Horava's theory
relaxes Lorentz invariance in the UV limit and is power-counting
renormalizable. In the IR limit, this kind of quantum gravity
reduces to the general relativity [50-52].

It is necessary to study the Casimir effect for parallel plates
within the frame of HL theory. The Casimir effect has been
analyzed widely in many areas of physics, so the effect can be
thought as a window to probe many kinds of models [8, 10, 12-15].
As mentioned above, the HL model provides the kinetic terms with
higher order spatial derivatives in the Lagrangian. This extension
will violate the Lorentz symmetry to change the dispersion
relation of original quantum field theory. The Lorentz-violating
extensions certainly revise the Casimir effect noticably. A lot of
efforts has been paid to the Casimir effect in the
Horava-Lifshitz-like theories. The experimental investigation on
micrometer-size physics has inspired the theoretical research on
the Casimir effect in the Lorentz-violating generalizations of the
standard quantum field theory whose Lorentz symmetry is fully
conserved [59-64]. The Casimir effect was considered in the
context of Lorentz-breaking scenarios [65-69]. The Casimir effect
for two parallel plates was studied in the anisotropic spacetime
dominated by HL proposal and the revised results of the effect
deviate from the measurements greatly [66]. The authors of Ref.
[67] discussed a massless scalar real field satisfying the
Dirichlet, Neumann and mixed boundary conditions respectively in
the HL-like theory to reflect the dependences of the Casimir
effects for parallel plates or two-dimensional rectangular boxes
on the HL parameter relating to the breaking of Lorentz symmetry
and the types of boundary conditions [67]. The Casimir energy of a
scalar field trapped in a box in the spacetimes with three spatial
Lifshitz dimensions was computed and the energy will be singular
or regular according to the value of the critical exponent
respectively [70]. In a box in the background with a Lifshitz
extra dimension, the Casimir energy was also evaluated and the
influence from Lifshitz parameter is manifest [70]. The Casimir
energy for parallel plates in a Lifshitz-like field theory was
considered and the relations among the Casimir energy, Lifshitz
fixed point parameter, mass of scalar field, dimensionality and
temperature were obtained [71].

It is significant to research on the Horava-Lifshitz theory
influence on the Casimir effect with nonzero temperature because
the quantum field theory at finite temperature shares many
effects. The thermal corrections to the Casimir energy under the
HL field theory around the Kehagias-Sfetsos black hole were found
[65, 72]. The Casimir free energy for parallel plates at a
Lifshitz fixed point (LP) was derived in the flat spacetime when
the temperature is not equal to zero in order to show the thermal
corrections to the Casimir effect and the results are important
[71]. It is important to scrutinize the Casimir effect at finite
temperature within the frame of HL scenario. We should further the
research in Ref.[71] to elaborate the Casimir energy and Casimir
force in terms of temperature in view of the modified gravity. We
should wonder how the temperature affects their natures. In order
to show the thermal corrections to the Casimir effect for parallel
plates at a Dirichlet boundary condition, we derive the frequency
of massless real scalar field with thermal corrections by means of
finite-temperature HL scheme. We regularize the frequency to
obtain the Casimir energy density with the help of the
hypergeometric functions and zeta function technique. The Casimir
force involving HL factor for parallel plates can also be obtained
from the Casimir energy per unit volume at finite temperature.
According to the expressions of the Casimir energy and Casimir
force, the discussions for the influence from HL theory and
temperature can be performed. The results will be listed in the
end.

\vspace{0.8cm} \noindent \textbf{II.\hspace{0.4cm}The Casimir
effect for parallel plates at finite temperature in the
Horava-Lifshitz theory}

We make use of the imaginary time formalism in the
finite-temperature field theory to describe the massless real
scalar field in thermal equilibrium [35]. A partition function for
the scalar field system introduces [35, 36, 50-52],

\begin{equation}
Z=N\int_{period}D\phi\exp[\int_{0}^{\beta}d\tau\int
d^{3}x\mathcal{L}(\phi, \partial_{E}\phi)]
\end{equation}

\noindent where $\mathcal{L}(\phi, \partial_{E}\phi)$ is the
Lagrangian density corrected by HL proposal. $N$ is a constant.
The subindex "period" lets the scalar field $\phi(\tau,
\textbf{x})$ to obey [35, 36],

\begin{equation}
\phi(0, \textbf{x})=\phi(\beta, \textbf{x})
\end{equation}

\noindent where $\tau=it$ and $\beta=\frac{1}{T}$, the inverse of
the temperature. The spatial coordinates are denoted as
\textbf{x}. Here the scalar field $\phi(\tau, \textbf{x})$
confined to the interior of parallel-plate device must satisfy the
Dirichlet boundary conditions at plates. According to the
solutions to the Klein-Gordon equations and the boundary
conditions, the generalized zeta function for massless real scalar
field can be written as [35, 36, 50-52, 66],

\begin{eqnarray}
\zeta(s; -\partial_{E})=Tr(-\partial_{E})^{-s}\hspace{5.5cm}\nonumber\\
=\int\frac{d^{2}k}{(2\pi)^{2}}\sum_{n=1}^{\infty}\sum_{l=-\infty}^{\infty}
[l^{2(z-1)}(k_{1}^{2}+k_{2}^{2}+(\frac{n\pi}{a})^{2})^{z}
+(\frac{2l\pi}{\beta})^{2}]^{-s}
\end{eqnarray}

\noindent where

\begin{equation}
\partial_{E}=\frac{\partial^{2}}{\partial\tau^{2}}-(-1)^{z}l_{0}^{2(z-1)}
(\nabla^{2})^{z}
\end{equation}

\noindent Here $k=\sqrt{k_{1}^{2}+k_{2}^{2}}$ denotes the
transverse components of the momentum. As mentioned above, $z$ is
the critical exponent [50-52]. $a$ is the distance of the plates.
We derive the generalized zeta function (3) to obtain,

\begin{eqnarray}
\zeta(s; -\partial_{E})\hspace{10cm}\nonumber\\
=\frac{1}{4\pi}\frac{1}{z s-1}(\frac{\pi}{a})^{-2s z+2}\zeta(2s
z-2)\hspace{6.5cm}\nonumber\\ +\frac{1}{2\pi}\frac{1}{z
s-1}l_{0}^{-2s(z-1)}\sum_{n_{1}=1}^{\infty}\sum_{n_{2}=1}^{\infty}
(\frac{n_{1}\pi}{a})^{-2s z+2}F(s, s-\frac{1}{z}, s-\frac{1}{z}+1;
-\beta_{1})
\end{eqnarray}

\noindent where

\begin{equation}
\beta_{1}=l_{0}^{-2(z-1)}(\frac{n_{1}\pi}{a})^{-2z}
(\frac{2n_{2}\pi}{\beta})^{2}
\end{equation}

\noindent and $F(a, b, c; x)$ is the hypergeometric function [73].
We obtain the total energy density of the system under thermal
influence [35, 36],

\begin{eqnarray}
\varepsilon=-\frac{\partial}{\partial\beta}(\frac{\partial\zeta(s;
-\partial_{E})}{\partial s}|_{s=0})\hspace{7cm}\nonumber\\
=\sum_{n=1}^{\infty}\frac{2\times(-1)^{n}}{nz+1}(\frac{1}{2\pi})^{2n+1}
l^{2(z-1)n}\zeta(-2(nz+1))\zeta(2n)(\frac{\pi}{a})^{2nz+2}\beta^{2n-1}\nonumber\\
+\frac{1}{\sin\frac{\pi}{z}}(1-\frac{1}{z})l^{-2(2-z-\frac{1}{z})}
(2\pi)^{2(\frac{1}{z}-1)}\zeta(-2z)\zeta(2(1-\frac{1}{z}))
()\frac{\pi}{a}^{2z}\beta^{1-\frac{2}{z}}\nonumber\\
-\frac{1}{2z\sin\frac{\pi}{z}}l^{-2(1-\frac{1}{z})}(2\pi)^{\frac{2}{z}}
\zeta(-\frac{2}{z})\beta^{-\frac{2}{z}-1}\hspace{4.5cm}\nonumber\\
+\frac{1}{\sin\frac{\pi}{z}}\sum_{n=1}^{\infty}(-1)^{n}l^{-2(z-1)(\frac{1}{z}-1-n)}
(2\pi)^{2(\frac{1}{z}-1-n)}(n+1-\frac{1}{z})\hspace{1cm}\nonumber\\
\times\zeta(-2(n+1)z)\zeta(2(n+1-\frac{1}{z}))(\frac{\pi}{a})^{2z+2nz}
\beta^{2n+1-\frac{2}{z}}\nonumber\\
+\frac{1}{\sin\frac{\pi}{z}}\sum_{n=1}^{\infty}(-1)^{n}l^{-2(z-1)(\frac{1}{z}-n)}
(2\pi)^{2(\frac{1}{z}-n)}(n-\frac{1}{z})\zeta(-2nz)\hspace{1cm}\nonumber\\
\times\zeta(2(n-\frac{1}{z}))(\frac{\pi}{a})^{2nz}\beta^{2(n-\frac{1}{z})-1}\hspace{3cm}
\end{eqnarray}

\noindent with the help of properties of hypergeometric function
and the Euler's reflection formula of Gamma function [73]. We
regularize the expression by means of zeta function technique like
$\zeta(1-s)=2(2\pi)^{-s}\Gamma(s)\zeta(s)\cos\frac{s\pi}{2}$ to
obtain the Casimir energy per unit volume for parallel plates at
finite temperature as follow,

\begin{eqnarray}
\varepsilon_{C}=\sum_{n=1}^{\infty}\frac{(-1)^{n}}{z
n+1}2^{-2(z+1)n-2}\pi^{-2\pi-2}l_{0}^{2(z-1)n}a^{-2z
n-2}\beta^{2n-1}\hspace{2cm}\nonumber\\
\times\Gamma(2z n+3)\zeta(2z n+3)\zeta(2n)\sin z
n\pi\hspace{2cm}\hspace{2cm}\nonumber\\
-\frac{z-1}{z\sin\frac{\pi}{z}}2^{-2z+\frac{2}{z}-2}\pi^{\frac{2}{z}-3}
l_{0}^{2(z+\frac{1}{z}-2)}a^{-2z}\beta^{1-\frac{2}{z}}\hspace{3.5cm}\nonumber\\
\times\Gamma(2z+1)\zeta(2z+1)\zeta(2-\frac{2}{z})\sin
z\pi\hspace{3cm}\nonumber\\
+\frac{1}{2\pi}l_{0}^{2(\frac{1}{z}-1)}\Gamma(\frac{2}{z}+1)\zeta(\frac{2}{z}+1)
\hspace{5.5cm}\nonumber\\
+\frac{1}{\sin\frac{\pi}{z}}\sum_{n=1}^{\infty}(-1)^{n}(n+1-\frac{1}{z})
2^{-2(z+1)n+\frac{2}{z}-4}\pi^{-2n+2z+\frac{2}{z}-5}\hspace{1cm}\nonumber\\
\times
l_{0}^{-2(z-1)(\frac{1}{z}-1-n)}a^{-2zn-2z}\beta^{2n+1-\frac{2}{z}}\hspace{2.5cm}\nonumber\\
\times\Gamma(2zn+3)\zeta(2zn+3)\zeta(2(n+1-\frac{1}{z}))\sin
zn\pi\nonumber\\
-\frac{1}{\sin\frac{\pi}{z}}\sum_{n=1}^{\infty}(-1)^{n}(n-\frac{1}{z})
2^{-2(z+1)n+\frac{2}{z}}\pi^{-2n+\frac{2}{z}-1}l_{0}^{-2(z-1)(\frac{1}{z}-n)}\hspace{0.5cm}\nonumber\\
\times
a^{-2zn}\beta^{2(n-\frac{1}{z})-1}\Gamma(2zn+1)\zeta(2zn+1)\zeta(2(n-\frac{1}{z}))\nonumber\\
\times\sin zn\pi\hspace{6.5cm}
\end{eqnarray}

\noindent The Casimir energy consists of five parts if the HL
variable is not chosen to be an integer. Four of the parts contain
$\sin z\pi$ or $\sin zn\pi$ terms, where $n$ is a positive
integer. If the critical exponent $z$ is equal to integer, the
Casimir energy will reduce to be,

\begin{equation}
\varepsilon_{C}\mid_{z=integer}=\frac{l_{0}^{2(\frac{1}{z}-1)}}{2\pi
z}\Gamma(1+\frac{2}{z})\zeta(1+\frac{2}{z})>0
\end{equation}

\noindent with the positive sign of the Casimir energy, no matter
the thermal influence proceeds. Within the frame of the HL model,
the Casimir effect for parallel plates can not exist due to the
positive due to the positive Casimir energy, which is not
consistent with the original experiment [3, 5-7]. According to
Eq.(9), it is obvious that the term left is independent of the
plate gap $a$. Further the Casimir force between the parallel
plates will disappear. We can adjust the parameter $z$ to simulate
the Casimir energy while we keep the critical exponent not to be
any integer, and the Casimir energy can be similar to the
measurements [3, 5-7].

The Casimir force between the plates is obtained by the derivative
of the Casimir energy (8) with respect to the plate distance. The
Casimir force per unit area with HL corrections on the plates
obeying the Dirichlet boundary conditions can be written as,

\begin{eqnarray}
f_{C}=-\frac{\partial\varepsilon_{C}}{\partial
a}\hspace{9cm}\nonumber\\
=\sum_{n=1}^{\infty}(-1)^{n}2^{-2(z+1)n-1}\pi^{-2n-2}l_{0}^{2(z-1)n}
a^{-2zn-3}\beta^{2n-1}\hspace{2.5cm}\nonumber\\
\times\Gamma(2zn+3)\zeta(2zn+3)\zeta(2n)\sin
zn\pi\hspace{3.5cm}\nonumber\\
-\frac{z-1}{\sin\frac{\pi}{z}}2^{-2z+\frac{2}{z}-1}\pi^{\frac{2}{z}-3}
l_{0}^{2(z+\frac{1}{z}-2)}a^{-2z-1}\beta^{1-\frac{2}{z}}\hspace{3.5cm}\nonumber\\
\times\Gamma(2z+1)\zeta(z+1)\zeta(2-\frac{2}{z})\sin
z\pi\hspace{4cm}\nonumber\\
+\frac{1}{\sin\frac{\pi}{z}}\sum_{n=1}^{\infty}(-1)^{n}(zn+1)(n+1-\frac{1}{z})
2^{-2(z+1)n+\frac{2}{z}-3}\pi^{-2n+2z+\frac{2}{z}-5}\nonumber\\
\times
l_{0}^{-2(z-1)(\frac{1}{z}-1-n)}a^{-2zn-2z-1}\beta^{2n+1-\frac{2}{z}}\hspace{2.5cm}\nonumber\\
\times\Gamma(2zn+3)\zeta(2zn+3)\zeta(2(n+1-\frac{1}{z}))\sin
zn\pi\nonumber\\
-\frac{1}{\sin\frac{\pi}{z}}\sum_{n=1}^{\infty}(-1)^{n}(z-n)2^{-2(z+1)n+\frac{2}{z}+1}
\pi^{-2n+\frac{2}{z}-1}\hspace{2.5cm}\nonumber\\
\times l_{0}^{-2(z-1)(\frac{1}{z}-n)}a^{-2zn-1}\beta^{2(n-\frac{1}{z})-1}\hspace{3cm}\nonumber\\
\times\Gamma(2zn+1)\zeta(2zn+1)\zeta(2(n-\frac{1}{z}))\sin
zn\pi\hspace{1cm}
\end{eqnarray}

\noindent According to Eq. (10), the exponents of the plates
distance such as $-(2zn+3)$, $-(2z+1)$, $-(2zn+2z+1)$ and
$-(2zn+1)$ are negative. When the parallel plates leave away from
each other, we discover that

\begin{equation}
\lim_{a\longrightarrow\infty}f_{C}=0
\end{equation}

\noindent which is favoured by the measurements. The Casimir force
per unit area at finite temperature was shown in Figure 1. The
shapes of the force curves with adequate value of HL factor is
similar to the standard Casimir force between the parallel plates.
We can regulate the HL parameter to correct the curves in
comparison with the experimental results. It is significant to
research on the Casimir force to estimate the HL exponent due to
the thermal influence. The features of the standard Casimir force
between the parallel plates [3, 5-7] can remain under HL
influence. It is manifest that the attractive Casimir force will
be weaker as the temperature increases. The curves of the force
with different temperatures are similar.

\vspace{0.8cm} \noindent \textbf{III.\hspace{0.4cm}Discussion}

We discuss the Casimir force for parallel plates under thermal
influence within the frame of Horava-Lifshitz field theory to
further the works on the Casimir free energy in Ref. [71]. We
derive the Casimir energy of the parallel-plate system which is
described at finite temperature under the HL issue. Further the
corresponding Casimir force is obtained. Within the
Horava-Lifshitz theory, the Casimir energy for parallel-plate
system is positive and the Casimir force between the two parallel
plates vanish, unless the HL factor $z$ is not equal to be an
integer. The magnitudes of the Casimir force between two parallel
plates with adequate HL parameter become small as the thermal
influence is stronger while the Casimir force remains attractive.
When the parallel plate distance is extremely great, the
asymptotic value of the Casimir force approaches to zero no matter
what the temperature is equal to while the terms involving the HL
exponent appearing in the expression. We argue that the HL
exponent from a kind of field theory can not be chosen as integers
exactly according to the results above. It is possible to select
the HL exponent to lead the Casimir effect of various models to be
comparable to the experimental results and we could explore the
special field theory in another direction in future.

\vspace{1cm}
\noindent \textbf{Acknowledge}

This work is partly supported by the Shanghai Key Laboratory of
Astrophysics.

\newpage
\begin{figure}
\setlength{\belowcaptionskip}{10pt} \centering
\includegraphics[width=15cm]{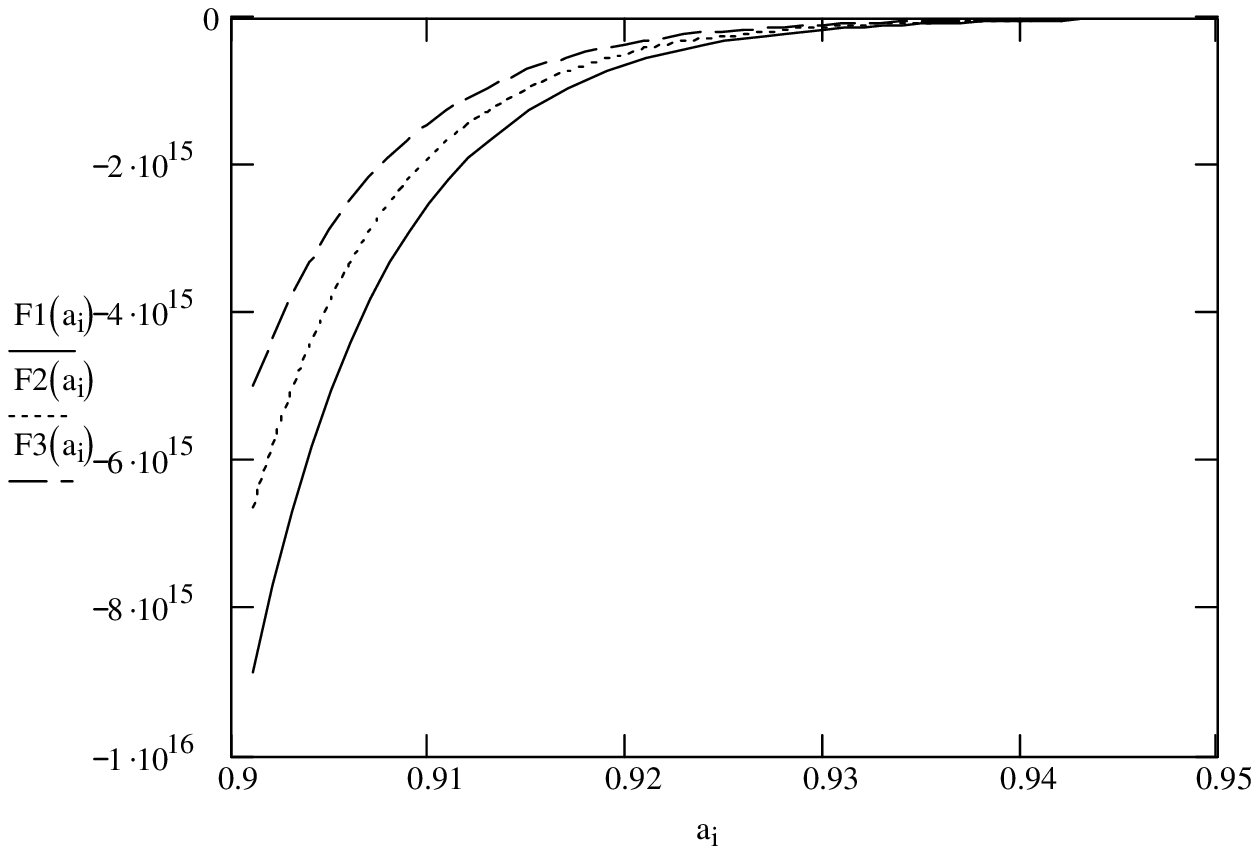}
\caption{The solid, dotted and dashed curves of the Casimir force
per unit area between two parallel plates as functions of plate
separation $a$ for $\beta=0.0101, 0.01005, 0.01$ respectively with
HL factor $z=2.1$.}
\end{figure}

\end{document}